# LANGUAGE MODELS AND A SECOND OPINION USE CASE: THE POCKET PROFESSIONAL

David Noever

PeopleTec, 4901-D Corporate Drive, Huntsville, AL, USA, 35805

david.noever@peopletec.com

## ABSTRACT

This research tests the role of Large Language Models (LLMs) as formal second opinion tools in professional decision-making, particularly focusing on complex medical cases where even experienced physicians seek peer consultation. The work analyzed 183 challenging medical cases from Medscape over a 20-month period (2023-2024), testing multiple LLMs' performance against crowd-sourced physician responses. A key finding was the high overall score possible in the latest foundational models (>80% accuracy compared to consensus opinion), which exceeds most human metrics reported on the same clinical cases (450 pages of patient profiles, test results and treatment recommendations). The study rates the LLMs' performance disparity between straightforward cases (>81% accuracy) and complex scenarios (43% accuracy), particularly in these cases generating substantial debate among human physicians. The study's focus on text-based responses does not fully attempt to capture the multimodal nature of medical diagnosis in these "human-hard" cases. However, the research demonstrates that LLMs may be valuable as generators of comprehensive differential diagnoses rather than as primary diagnostic tools, potentially helping to counter cognitive biases in clinical decision-making, reduce cognitive loads and thus remove some sources of medical error. The inclusion of a second comparative legal dataset (Supreme Court cases, N=21) provides added empirical context to the AI use to foster second opinions, though these legal challenges proved considerably easier for LLMs to analyze. In addition to the original contributions of empirical evidence for LLM accuracy, the research aggregated a novel benchmark for others to score highly contested question and answer reliability between both LLMs and disagreeing human practitioners. These results suggest that the optimal deployment of LLMs in professional settings may differ substantially from current approaches that emphasize automation of routine tasks.

## KEYWORDS

*Generative AI, Transformers, Professional Certification, Large Language Models*

## 1. INTRODUCTION

Recent workers have integrated Large Language Models (LLMs) into professional workflows, primarily as assistants for routine tasks such as documentation, summarization, and information retrieval [1-15]. While LLMs have performed well on medical licensing exams, this passing grade may mask the complexity gap between test-taking and real-world clinical reasoning. The structured nature of board examinations, with their carefully curated answer spaces and well-defined problem boundaries, presents a fundamentally different challenge from the open-ended, nuanced differential diagnoses encountered in daily medical practice. This paper explores this distinction experimentally by examining cases where even experienced physicians seek peer consultation through professional forums like Medscape—cases that often defy traditional diagnostic frameworks and challenge the boundaries of medical consensus. We extend the second-opinion framework to a Supreme Court legal dataset for generality and comparison. The original contributions of the work therefore include the empirical assessment of the second-opinion paradigm for AI utility along with two new human-rated professional datasets that require reasoning nuance beyond traditional licensing examinations.

Our research focuses on these "human-hard" cases: clinical scenarios where the pathway to diagnosis remains ambiguous enough to prompt practicing physicians to seek crowdsourced wisdom. By analyzing hundreds of such cases from professional medical forums alongside LLM responses to the same scenarios, we test both the potential and limitations of artificial intelligence in navigating medical uncertainty. The results suggest that while LLMs may not replace human clinical judgment, they could serve a unique role in augmenting decision-making processes, particularly in gray areas where multiple interpretations of available evidence exist.

This investigation challenges the current paradigm of positioning LLMs primarily as automation tools for routine medical tasks. Instead, we propose a framework for deploying these systems as specialized agents for second opinions, particularly in scenarios where traditional diagnostic approaches have reached their limits. Our findings suggest that the true value of LLMs in medicine may lie not in their ability to replicate standard medical knowledge, but in their capacity to systematically explore the gaps and uncertainties that characterize real-world clinical practice.

The literature on large language models (LLMs) in healthcare has primarily focused on evaluating their performance as diagnostic tools, educational aids, and participants in medical exams. Studies have assessed ChatGPT's diagnostic utility for both clinicians and students, as well as its effectiveness compared to traditional online symptom checkers and triage apps, providing a foundation for understanding its role in clinical settings [1-4]. A significant body of research has explored ChatGPT's ability to pass medical licensing and specialty exams across various countries and domains, including China, Japan, the United States, and the United Kingdom [5-17]. These evaluations highlight both the potential and limitations of LLMs, emphasizing variability across different models and exam types [7-12].

Recent systematic reviews and meta-analyses consolidate findings from multiple studies, providing insights into the strengths and weaknesses of LLMs across diverse testing scenarios [7, 18-21, 28]. While ChatGPT has shown some success in exams like family medicine [22], it has also encountered notable challenges, such as failing Taiwan's family medicine board exam [17]. Moreover, discussions on LLM performance have expanded to include ensemble learning approaches, which suggest that combining multiple models may enhance reasoning capabilities and reliability in medical question answering [24-25].

LLMs are increasingly viewed not as standalone diagnostic agents but as complementary tools that alleviate cognitive load and bridge knowledge gaps. This application reflects the shift toward integrating LLMs into collaborative workflows, where they act as advisory systems rather than autonomous decision-makers [26-27]. The current study builds on this evolving perspective by examining how LLMs function as second-opinion guides, distinct from fully autonomous agents, thereby focusing on their ability to assist medical professionals without replacing clinical oversight in dataset development [29-31]. This approach aligns with recent trends, aiming to enhance the decision-making process through AI-supported collaboration rather than substitution.

## 2. METHODS

Our study analyzed case challenges from Medscape's professional forum [29,31] over a 20-month period from January 2023 through October 2024. This timeframe yielded real-world 183 challenging cases, representing approximately one case every three days. Each case consisted of detailed clinical presentations submitted by practicing physicians seeking diagnostic consultation, with an average length of several pages of text. Cases included two multiple-choice questions with corresponding crowd-sourced responses from the medical community [31]. This specific timeframe was selected to address potential training data contamination concerns, as recent studies have demonstrated performance degradation in LLMs when evaluating medical examination questions outside their training windows.

In Figure 1, an example mind map organizes the two-page clinical case based on key features such as patient profile (demographics and background), symptoms (complaints, timing, characteristics), physical exam findings (deficits and anomalies), diagnostic tests (normal and abnormal results), differential diagnosis (explanation of alternatives), and treatment plan (medication and therapy). While a traditional medical textbook might highlight how a qualified medical practitioner would deduce the best treatment course, the primary point here is to illustrate the multitude of decisions and wrong turns that a rule-based AI might encounter. The final note centers on whether a consulting LLM suggests reasoning rather than just language manipulation when traversing a life and death decision.

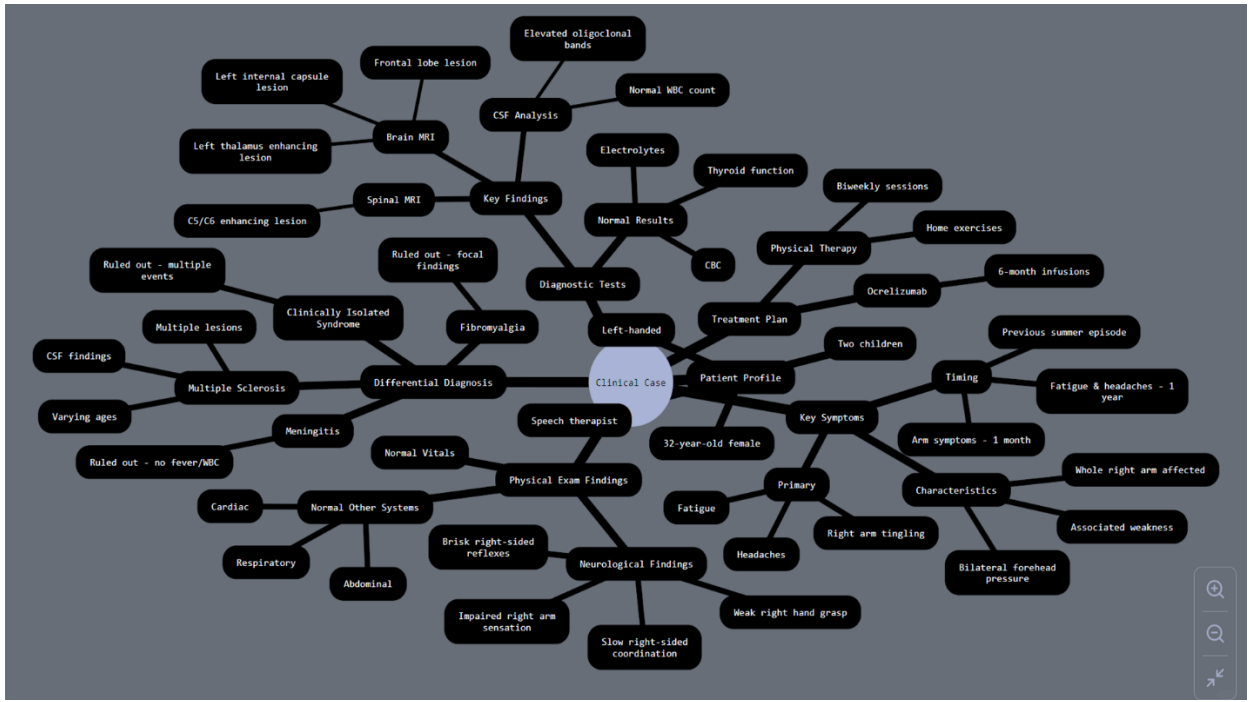

*Figure 1. A mind map created to analyze the complexity of a single clinical case file in Medscape challenges (prompted to Claude and using MermaidJS mind map feature based on 2.2 pages of clinical observations)*

In addition to highlighting complexities, Figure 1 also gauges the cognitive load on a human doctor to complete these tough cases by estimating the test environment of a licensing exam. The selected 183 cases span 2.2 pages per example not including the questions and answers. The total test comprises 450 pages (317k words). The reading (Flesch-Kincaid) grade level of the exposition is college (15.2 grade) with reading ease of 22.6. These measures weight the syllables per words and sentence lengths which in technical writing are typically large (e.g. 22 words/sentence). A human doctor reading at approximately five words per second would take 17 hours in a sustained reading session to finish mechanically the clinical narrative.

| Specialization | Field | Score |
|---|---|---|
| Allergy and Immunology | 1% | 100% |
| Diagnostic Radiology | 0% | 100% |
| Emergency Medicine | 2% | 100% |
| Family Medicine | 0% | 100% |
| Internal Medicine | 1% | 100% |
| Neurology | 3% | 100% |
| Pediatrics | 1% | 100% |
| Preventive Medicine | 0% | 100% |
| Surgery | 1% | 100% |
| Urology | 1% | 100% |
| Psychiatry | 2% | 89% |
| General Medicine | 82% | 79% |
| Dermatology | 5% | 78% |
| Pathology | 1% | 75% |
| **Grand Total** | **361** | **293** |

*Figure 2. Questions by Specialty in Medscape Challenges and Corresponding Best-Case Scores for GPT-4o.*

By medical specialties, 82% of the Medscape challenges were categorized (by ChatGPT) as General Medicine cases, with dermatology as the second specialty (5%). Figure 2 shows the distribution of challenge cases by sub-specialty as assigned by GPT-4o from the menu of choices and case history descriptions. The dermatology, pathology, and radiology cases featured more image supporting documentation that was not included in the question-answer phases. To emphasize this ambiguity, the average number of answers with significant doctor selected votes was 2.14 (out of 4 possibilities), thus making the differential diagnoses at least a coin-flip.

The evaluation protocol included closed-source foundation models (Google, Open AI, Anthropic), open-source models (Meta, Alibaba, Microsoft), and vision-capable multimodal models (LLaVa). For standardization, all initial evaluations were conducted using text-only case histories, excluding diagnostic images. This approach ensured comparability across all models, including those without vision capabilities. For cases that included diagnostic imaging, we conducted additional comparative analysis using vision-capable models to evaluate the impact of multimodal information on diagnostic accuracy. All model interactions were conducted via API calls to ensure no context retention between queries, consistent evaluation conditions, and elimination of potential chat interface artifacts. Our query structure prompted each model to provide its primary diagnostic choice, detailed reasoning, alternative diagnoses considered, and key discriminating factors between primary and alternative diagnoses.

As shown in Figures 3-4, performance evaluation focused on agreement with crowd-sourced physician consensus, quality, and depth of supporting reasoning, identification of diagnostic uncertainties, recognition of cases requiring additional information, and appropriate citation of clinical guidelines. For cases with imaging, we specifically analyzed the impact of visual information on diagnostic accuracy only by specialties (dermatology, radiology, and pathology). Statistical analysis employed standard measures of inter-rater agreement between model outputs and physician consensus, with particular attention to cases where significant disagreement occurred. In this study, several metrics were used to analyze the spread (standard deviation) and consensus within physicians' responses. Entropy served as a key measure of response dispersion, with higher values indicating a broader distribution of answers, suggesting a lack of agreement among participants. The Top2 Ratio captured the relationship between the most frequent and second most frequent responses, with higher ratios reflecting stronger consensus as the leading answer clearly dominated the distribution. Additionally, the Max Percent metric identified the highest percentage for any single response, highlighting the extent to which one answer stood out.

| Model | Score (%) |
|---|---|
| Open AI GPT-4-turbo | 81.2% |
| Open AI GPT-4o | 81.2% |
| Open AI ChatGPT-4o-latest | 80.3% |
| Anthropic Claude-3-Opus-20240229 | 80.0% |
| Anthropic Claude-3.5-Sonnet-20241022 | 78.7% |
| Open AI GPT-4o-mini | 78.4% |
| Anthropic Claude-3.5-Sonnet-20240620 | 74.2% |
| Anthropic Claude-3-Haiku-20240307 | 70.4% |
| Google Gemini-1.5-Flash | 64.0% |
| Anthropic Claude-2.0 | 62.9% |
| Open AI GPT-3.5-turbo | 59.5% |
| Meta LLaMa-3-latest | 54.9% |
| Google Gemma2-latest | 54.3% |
| Meta LLaMA-3.1-latest | 51.5% |
| Mistral-latest | 45.1% |
| LLaVa-latest | 44.9% |
| Dolphin-Mistral-latest | 38.0% |
| LLaMa-2-Uncensored-latest | 7.8% |

*Figure 3. Medscape Physician Challenge Results (OCT 2024 models). 23 models were tested, 5 were not instructible in answer format requirements.*

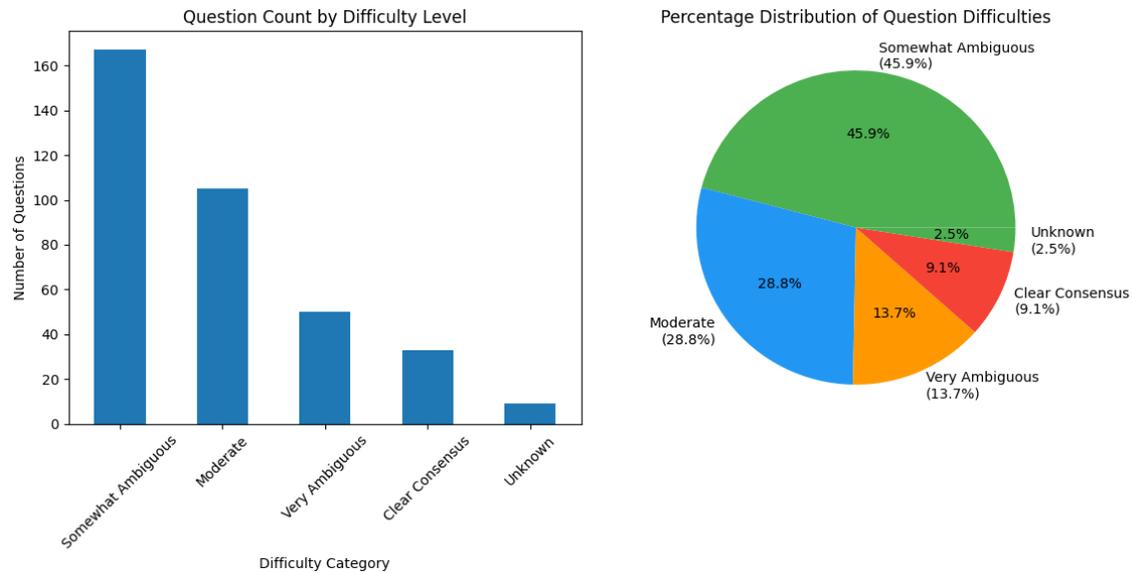

*Figure 4. Challenge Case Difficulty based on Crowd-sourced Physician Opinions*

To interpret these metrics, specific ranges of entropy were used to classify the level of ambiguity in the responses. Values above 1.2 indicated a highly ambiguous situation where physicians' answers were widely dispersed, while entropy between 0.8 and 1.2 suggested moderate uncertainty. When entropy fell between 0.4 and 0.8, it reflected moderate ambiguity with some emerging consensus. Finally, entropy values below 0.4 signaled clear agreement among students, demonstrating a strong alignment toward a particular answer. These metrics provided a comprehensive view of the variability in physician responses, allowing for the identification of areas that may benefit from further clarification or instructional support.

## 3. RESULTS

The challenge cases showed significant spread among human physicians in the test itself, prior to including LLM choices. Applying the entropy (information content) metric for the opinion differences (spread), 74.7% (272/364 case questions) ranked as either "somewhat ambiguous" (45.9%) or "moderately ambiguous" (28%). The "very ambiguous" (13.7%) compared roughly equivalent to the "clear consensus" (10.7%). This spread in opinion should not surprise since the challenge cases are selected for their difficulty to a community practicing differential diagnosis. No notation in Medscape challenges indicate whether a particular specialist opinion might hold more weight in reported answers (although that pathway dominates the likely real-world scenarios where a difficult case travels from a generalist to a specialist).

Analysis of the 183 case challenges and 361 (viable) questions revealed significant disparities between LLM performance on straightforward diagnostic scenarios versus complex cases that prompted physician forum consultation in Figures 4-5. While foundational large models demonstrated high accuracy (>81%) in matching physician consensus on cases with clear diagnostic criteria, performance declined markedly (43%) in scenarios characterized by atypical presentations or competing diagnostic possibilities in Figure 5. Notably, cases that generated substantial debate in physician forums showed the lowest LLM consensus matching (44%). The study defines disagreement as <1.2 ratio between first and second consensus answer as its metric for disagreement in crowd-sourced responses which represent the most difficult 25 human-hard questions.

The inclusion of detailed reasoning alongside diagnostic choices revealed important patterns in model behavior. When correct, models typically provided explanations closely aligned with established clinical decision trees. However, in cases where models diverged from physician consensus, their reasoning often failed to account for subtle clinical nuances that forum participants highlighted as crucial discriminating factors. Multiple models showed patterns of guessing or an inability to follow instructions. The smaller open-source models (Microsoft Phi-3.5, Zephyr AI Zephyr, Alibaba Gwen-2.5, Meta LLaMa-2) were deemed impossible to score in this test because they failed to pick multiple-choice answers and instead opted for summaries of the challenge cases. This pattern was particularly evident in cases involving multisystem disorders or atypical (often image-based) presentations of common conditions, where human experts explicitly cited pattern recognition from clinical experience as key to their diagnostic reasoning. Meta LLaMa-3-latest was an example of the model scoring its own uncertainties by either answering outside the scope (with unavailable answer choices-E,P,T,M) or simply default choosing "A" as an option.

In Figure 2, the best model comparison between text-only and multimodal models on the subset of cases with either dermatological or pathological imaging (n=24) demonstrated text limitations in diagnostic accuracy without visual data access (76% vs. 81% consensus matching). However, this improvement was not uniform across all studied cases and sub-specialties (with exception for radiological clinical evidence).

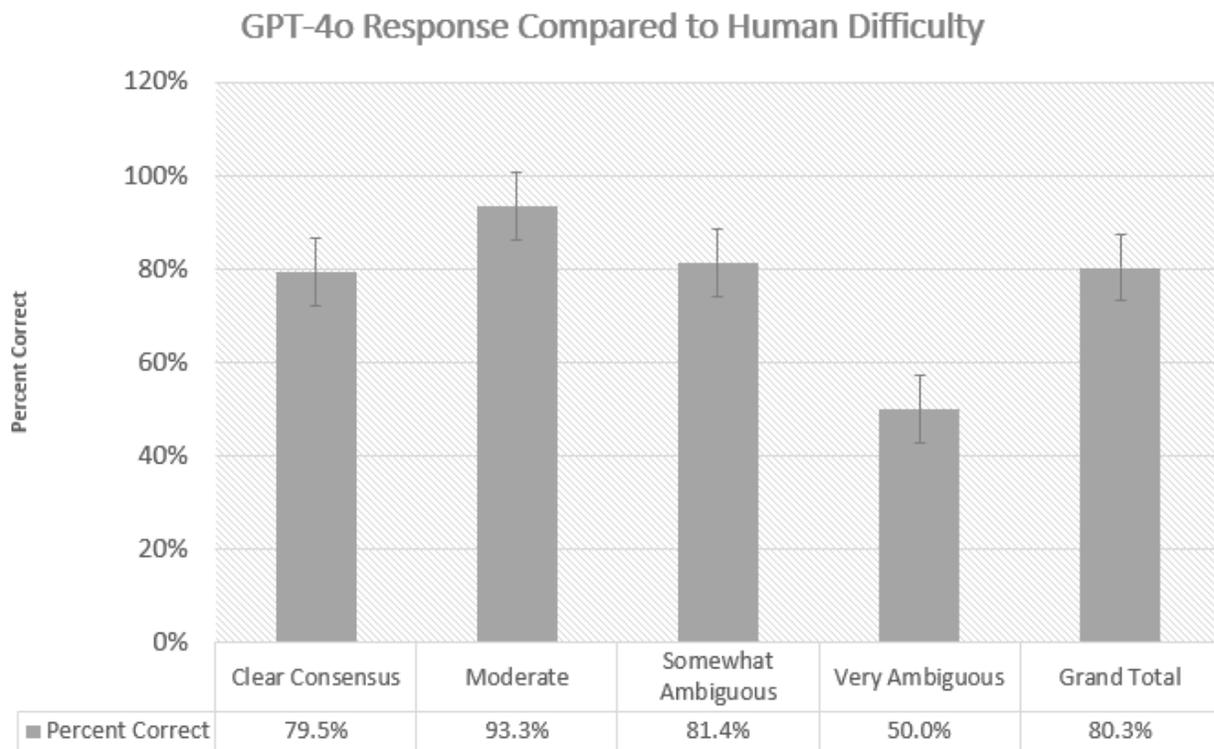

*Figure 5. GPT-4o Correctness Compared to Human Difficulty in Practicing Physicians*

Cross-model comparisons revealed consistency in areas of LLMs' difficulty, with all models struggling similarly with cases that physicians labeled as requiring "clinical gestalt" or "pattern recognition from experience." Interestingly, models frequently identified appropriate alternative diagnoses in their reasoning, even when their primary diagnostic choice differed from physician consensus. This suggests potential utility in second-opinion scenarios, where comprehensive differential diagnosis generation is valuable.

For generality, we tested legal knowledge using Supreme Court cases with variable human disagreement based on the published split [30] in the judicial decisions. This much smaller dataset (N=21) followed the same multiple-choice format and proved much easier for even the smaller models like Microsoft Phi-3.5 to get perfect scores. Figure 7 summarizes the per model score on this legal second opinion examination. That dataset is also available for review [31] but proved considerably easier for the current generation of LLMs and not as to examine the role of disagreement as a measure of "human-hard" decision-making by AI. Because of the historical range, one can assume the main reason for this relative ease stems from the court cases appearing in training data in contrast to the most recent medical cases which necessarily represent true validation runs outside the available update cycle of current foundational LLMs.

## 4. DISCUSSION AND FUTURE WORK

Our findings measure LLMs' ability to process medical information and their capacity to replicate the nuanced decision-making of experienced clinicians. While previous research has emphasized these models' success on standardized medical examinations, our analysis of real-world clinical challenges reveals potential uses in their diagnostic reasoning, particularly in cases that human physicians find sufficiently complex to warrant peer consultation.

The consistent performance degradation across all models in cases requiring ambiguity points to a fundamental limitation in current language model architectures. However, this limitation reveals an unexpected strength: the models' capacity to generate expansive differential diagnoses often exceeded the scope considered by individual human clinicians. This synthetic breadth suggests parallels with ongoing debates about AI creativity and invention, where the ability to systematically explore possibility spaces may compensate for lack of intuitive understanding. Just as these systems can generate novel technical solutions by exhaustively exploring design spaces, they appear capable of surfacing diagnostic possibilities that might be overlooked by human practitioners constrained by cognitive limitations or practice patterns.

Particularly significant is the models' immunity to common cognitive biases that affect human clinical reasoning. Unlike human clinicians who may be unduly influenced by recent cases (recency bias) or seek information that confirms initial impressions (confirmation bias), LLMs consistently generated comprehensive differential diagnoses regardless of context. We removed model memory in a chat context by prompting each question independently within the vendor's API. This systematic approach to diagnostic consideration, while lacking the intuitive depth of human expertise, offers a valuable counterbalance to the psychological factors that can compromise clinical decision-making. The reduction in cognitive load afforded by automated differential diagnosis generation may allow clinicians to focus their expertise on evaluation and integration rather than initial possibility generation. While a human might spend 2-3 days answering 361 challenging questions, the LLMs can complete the same test in less than an hour.

| Model | Score (%) |
|---|---|
| claude-3-5-sonnet-20240620 | 100 |
| claude-3-5-sonnet-20241022 | 100 |
| claude-3-haiku-20240307 | 100 |
| claude-3-opus-20240229 | 100 |
| claude_21 | 100 |
| chatgpt-4o-latest | 100 |
| dolphin-mistral_latest | 100 |
| gemini-1.5-flash | 100 |
| gemma2_latest | 100 |
| gpt-3.5-turbo | 100 |
| gpt-4 | 100 |
| gpt-4o-mini | 100 |
| gpt-4o | 100 |
| phi3_5_latest | 100 |
| qwen2_5_latest | 100 |
| zephyr_latest | 100 |
| llama3_latest | 95 |
| mistral_latest | 95 |
| llama3_1_latest | 38 |
| llava_latest | 38 |
| llama2-uncensored_latest | 14 |
| llama2_latest | 0 |

*Figure 7. Supreme Court Legal Disagreement Scores based on Model Vendor and Size*

The light correlation between model performance and physician ambiguity raises interesting "second opinion" considerations for clinical implementation. Unlike human physicians, who often express uncertainty in complex cases and seek consultation, current models lack reliable internal calibration of their diagnostic confidence. This disconnect could pose risks in clinical settings where model outputs might be mistaken for authoritative rather than advisory input. But when human professionals gauge their doubt, they typically defer a patient to specialists or seek a second opinion. The role for a LLM consultant in this instance thus offers a novel approach to the copilot or personal assistant. Thus, this inability to respond with "I don't know" as a limitation also suggests a natural role for these systems in augmenting rather than replacing human decision-making, particularly in generating difficult or expanded differential diagnoses for consideration by human clinicians.

Our results suggest that the optimal deployment of LLMs in clinical settings may differ substantially from current approaches that emphasize automation of routine tasks. Instead, these systems may prove most valuable in their capacity to serve as systematic, tireless generators of diagnostic possibilities, particularly in complex cases where broadening the consideration set could help prevent diagnostic anchoring or premature closure. This role as agents of second opinion rather than primary diagnosis aligns more closely with both the demonstrated strengths and limitations of current language model architectures, while potentially reducing the impact of cognitive biases that can affect human clinical reasoning. This role also extends current medical testing of LLMs beyond simple licensing or certification stages and opens up inventive supplemental advice when even the most dedicated professionals might disagree.

Further investigation is needed to explore several key areas emerging from our findings. First, the development of specialized prompting strategies that better elicit clinically relevant pattern recognition warrants investigation, particularly approaches that might help bridge the gap between algorithmic and experiential knowledge. Additionally, research into methods for improving model confidence calibration could enhance their utility in clinical settings. For example, we did not attempt to do few-shot learning where a worked example can guide a LLM to a solution better than zero-shot learning where the prompt presents the question cold without additional hints or templated guidance. A longitudinal study tracking how model performance evolves with the integration of more recent medical literature and case studies could provide insights into the role of training data temporality in clinical reasoning capabilities. Retrieval-augmented generation (RAG) models specialize diagnostic hints into the pre-prompting of many expert systems.

The interaction between human clinicians and LLM-generated second opinions represents another crucial area for investigation. Studies examining how physicians integrate model-generated differential diagnoses into their clinical reasoning process could inform more effective implementation strategies. Furthermore, research into the potential of these systems to serve as educational tools for medical trainees, particularly in developing systematic approaches to differential diagnosis, could reveal additional valuable applications.

A targeted future research priority lies in quantifying the cognitive load reduction afforded by LLM-assisted differential diagnosis. As noted, a human reader could not absorb the clinical text except over multiple exhausting 8-hour days of continuous and relentless examination. The LLM on the other hand can respond in 10-100 times less time and thus trigger the human reader to confirm rather than decide the case verdicts. The cognitive load could further be quantified through a controlled study measuring physicians' cognitive resources using established metrics such as NASA Task Load Index (TLX) scores, eye-tracking patterns, and physiological stress indicators when approaching complex cases with and without LLM support. Time-motion analysis could track the allocation of cognitive resources across different diagnostic tasks, while standardized documentation of diagnostic pathways could reveal whether access to LLM-generated differential diagnoses allows physicians to devote more attention to critical thinking and case synthesis rather than initial hypothesis generation. Complementary qualitative research using structured interviews and think-aloud protocols could illuminate how clinicians integrate LLM suggestions into their cognitive workflow. Of particular interest would be measuring decision fatigue across sequential cases, testing the hypothesis that LLM assistance might help maintain consistent diagnostic thoroughness throughout long clinical shifts. Additionally, investigating the relationship between cognitive load reduction and diagnostic accuracy could help establish optimal integration points for LLM assistance in clinical workflows, potentially identifying specific case types or clinical scenarios where automated differential generation provides maximum benefit with minimal disruption to established clinical reasoning patterns.

## 5. CONCLUSIONS

Our analysis of LLM performance on challenging clinical cases reveals both important limitations and promising opportunities in the application of these technologies to medical diagnosis. While current models fall short of replicating the nuanced clinical reasoning of experienced physicians, they demonstrate potential value as systematic generators of diagnostic possibilities, particularly in complex cases where broadening the differential diagnosis could be beneficial.

The clear distinction between model performance on standardized examinations versus real-world clinical challenges suggests that current evaluation metrics for medical AI systems may need recalibration. Rather than pursuing direct replacement of human clinical judgment, our findings support a more nuanced role for these technologies as augmentative tools in the diagnostic process. The identified limitations in pattern recognition and experiential knowledge integration, rather than disqualifying these systems from clinical use, help define their most appropriate role: as agents of second opinion rather than primary diagnostic tools.

These results suggest that the near-term future of medical AI may lie not in autonomous diagnosis but in the integration of machine-generated insights into human clinical reasoning processes. By positioning LLMs as partners in diagnostic consideration rather than arbiters of diagnostic truth, we may better leverage their capabilities while acknowledging their limitations. This approach aligns with the fundamental goal of medical technology: not to replace human judgment, but to enhance it in service of improved patient care.

## ACKNOWLEDGMENTS
The author thanks the PeopleTec Technical Fellows program for encouragement and project assistance.